\documentclass[aps,pra,twocolumn,superscriptaddress,amsmath,amssymb]{revtex4}
\usepackage{graphicx,graphics}
\usepackage{amsmath,sidecap}
\usepackage{color}
\begin{document}

\title{Simulation and detection of Weyl fermions in ultracold Fermi gases \\ with Raman-assisted spin-orbit coupling}
\author{Cheng-Gong Liang}
\affiliation{Department of Physics and Electronic Engineering, Jinzhong University, Jinzhong 030619, China}
\author{Ze-Gang Liu}
\affiliation{CAS Key Laboratory of Quantum Information, University of Science and Technology of China, Hefei 230026, China}
\author{Wei Han\footnote{hanwei.irain@gmail.com}}
\affiliation{State Key Laboratory of Quantum Optics and Quantum Optics Devices, Institute of Opto-Electronics, Shanxi University, Taiyuan 030006, China}

\begin{abstract}
Weyl fermion, also referred to as pseudo-magnetic monopole in momentum space, is an undiscovered massless elementary particle with half-integer spin predicted according to relativistic quantum field theory. Motivated by the recent experimental observation of Weyl semimetal band in ultracold Bose gases with Raman-assisted 3D spin-orbit coupling, we investigate the properties and possible observation of Weyl fermions in the low-energy quasi-particle excitations of ultracold Fermi gases. Following a previous suggestion that the existing Raman lattice scheme can be readily generalized to fermionic systems, here we discuss the movement of the Weyl points in the Brillouin Zone, as well as the creation and annihilation of Weyl fermions by adjusting the effective Zeeman field. The relevant topological properties are also demonstrated by calculating the Chern number. Furthermore, we propose how to experimentally verify the existence of the Weyl fermions and the associated quantum phase transition via density profile measurements.
\end{abstract}

\maketitle

\section{Introduction}
Weyl fermions were initially predicted in particle physics by H. Weyl in 1929~\cite{HWeyl}. According to the Dirac equation, a massless electron can be distinguished as right-handed or left-handed chirality depending on propagating parallel or antiparallel to its spin. This massless electron with chirality is known as Weyl fermion. For almost a century, the scientists have been trying to directly detect this kind of relativistic particle, but no significant progress has been made in the field of particle physics.

In recent years, Weyl fermion has triggered great interests in other physical fields, involving condensed matter physics, optics and cold atom physics. It has been theoretically predicted~\cite{XWan,HWeng,MZHasan} and experimentally verified~\cite{SYXu,BQLv,XHuang} that the low-energy quasi-particle excitations near the band-crossing (degenerate) points of some topological semimetals resemble the well-known Weyl fermions. These topological semimetals with Weyl points have excellent topological properties~\cite{NArmitage}, and may contribute to the development of low-consumption topological electronics and highly fault-tolerant topological quantum computation. Almost at the same time, Weyl points was also observed in photonic crystals~\cite{LLu,LLu2015}, facilitating the development of high dimensional topological photonics~\cite{LLu2014,TOzawa}.

Considering the uncontrollability of parameters and the challenges in growing materials of topological semimetals, it is also highly desirable to simulate Weyl fermions with cold atoms. While several theoretical proposals using atom-light interactions was suggested~\cite{JHJiang,YXu,BLiu,TDubcek,DWZhang,SGaneshan,WYHe,ZLi,YXu2,XKong,KShastri}, the Raman lattice scheme of 3D spin-orbit coupling brought the first experimental realization of Weyl semimetal band in ultracold Bose atomic gases~\cite{SChen2021,XJLiu2016,XJLiu2020}. The advantages of the cold atom platform are the highly controllable and impurity free, which produce an ideal Weyl semimetal with the minimum number of Weyl cones being two. This opens a broad avenue for the simulation of Weyl physics. Although the existing experiment is focused on Bose gases, it is promising to generalize the Raman lattice scheme to fermionic systems, which may lead to the realization of abundant topological quantum phases and various relativistic quasiparticles. While the Weyl points have been experimentally measured via the virtual slicing imaging technique in Bose gases~\cite{SChen2021}, it is also interesting to consider how to observe the Weyl points in Fermi gases.
\begin{figure*}[tbp]
\centerline{\includegraphics[width=0.95\textwidth,clip=]{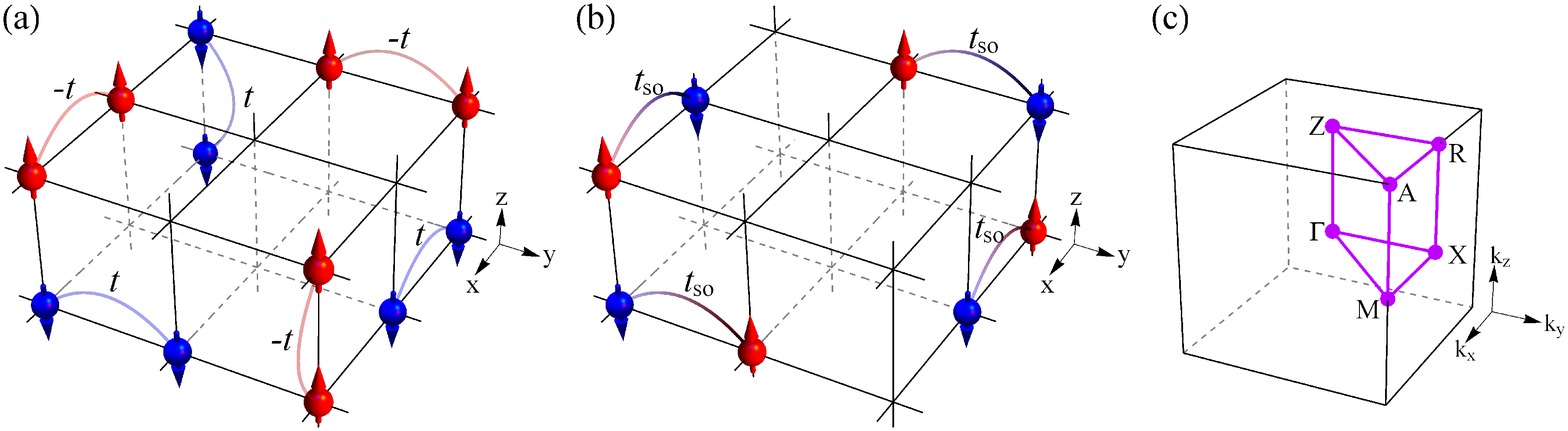}}
\caption{Cold atoms in the cubic optical lattice with Raman-assisted spin-orbit coupling. (a)-(b) Sketch of the cubic optical lattice with spin-conserved and spin-flip hopping denoted by $t$ and $t_{\mathrm{so}}$, respectively. (c) The Brillouin zone of the cubic optical lattice. The path $\Gamma$--$X$--$M$--$\Gamma$--$Z$--$R$--$A$--$Z|X$--$R|M$--$A$ is chosen in representing the energy band structures of Figs.~2 and~3. The paths $\Gamma$--$Z|X$--$R|M$--$A$ and $\Gamma$--$Z'|X$--$R'|M$--$A$ are chosen to track the trajectory of the Weyl points in Fig.~4(a).}\label{fig1}
\end{figure*}

Here we focus on the simulation and detection of Weyl fermions in ultracold Fermi gases with Raman-assisted spin-orbit coupling. We first study in detail the properties of the Weyl points in the Brillouin zone. Particularly, we focus on the movement of the Weyl points in the momentum space, as well as the creation and annihilation of Weyl fermions in the low-energy quasiparticle excitations. It is found that different number of Weyl points (tow or four) can emerge in the Brillouin zone depending on the strength of the effective Zeeman field. The Weyl points are always generated in pairs with opposite chirality, and the Berry curvature around the Weyl points clearly demonstrates that they resemble magnetic monopoles with a positive and negative magnetic charge, respectively. The movement trajectories of the Weyl points as a function of the strength of the Zeeman field are plotted, which exactly distinguish different topological phases. In some specific strengths of Zeeman field, the Weyl fermions annihilate in pairs at the high-symmetry points, and replaced with 2D massless Dirac fermions. For sufficiently strong Zeeman field, the band gap is opened and the low-energy excitations near the minimum gap manifest massive 3D Dirac fermions. Through explicit calculations, we show that the Weyl fermions and the associated phase transition can be experimentally observed by the conventional atomic detection techniques based on the density profile measurement~\cite{JRAnglin,MWZwierlein,YShin}.

The rest of the paper is organized as follows. In Sec.~\uppercase\expandafter{\romannumeral2}, we introduce the model Hamiltonian of ultracold Fermi atoms trapped in a cubic optical lattice with Raman-assisted 3D spin-orbit coupling. In Sec. \uppercase\expandafter{\romannumeral3}, by diagonalizing the Bloch Hamiltonian, we investigate the band structure as a function of the effective Zeeman field, and indicate the explicit symmetry broken of the band induced by spin-orbit coupling. In Sec. \uppercase\expandafter{\romannumeral4}, we show the distribution and movement of the Weyl points in the Brillouin zone by adjusting the strength of the effective Zeeman fields, and discuss the creation and annihilation of Weyl fermions in the low-energy quasi-particle excitations. The topological phase diagram is plotted according to the trajectories of the Weyl points. In Sec. \uppercase\expandafter{\romannumeral5}, we develop the method of density profile measurement to observe the Weyl fermions as well as the transformation between different relativistic quasi-particles. Finally, we summarize and give concluding remarks in Sec. \uppercase\expandafter{\romannumeral6}.

\section{Model}
We consider ultracold Fermi atoms trapped in a conventional cubic optical lattice with Raman-assisted 3D spin-orbit coupling~\cite{XJLiu2016,XJLiu2020}. The Hamiltonian is written as
\begin{eqnarray}
H&=&\left[\frac{\hbar^{2}\mathbf{k}^{2}}{2m}+V_{\mathrm{latt}}\left(x,y,z\right)\right]\otimes\mathbf{1+}
\mathcal{M}_{x}\left( x,y,z\right) \sigma _{x}\notag \\
&& +\mathcal{M}_{y}\left(x,y,z\right) \sigma _{y}+h_{z}\sigma _{z},\label{Hamiltonian}
\end{eqnarray}
where $\hbar$ is the Plank's constant $h$ divided by $2\pi$, $\mathbf{k}$ is the wave vector that represents the momentum of the atoms, $\mathbf{1}$ is the $2$-by-$2$ unit matrix, $\sigma_{x,y,z}$ are the Pauli matrices acting on the spins, $m$ is the mass of an atom, and $h_{z}$ is a tunable Zeeman constant. The spin-independent cubic lattice potential reads
\begin{equation}
 V_{\mathrm{latt}}=V_0[\cos^2(k_0x)+\cos^2(k_0y)+\cos^2(k_0z)],
\end{equation}
and the Raman coupling induced periodic potentials are given by
\begin{subequations}
\label{Blochvector}
\begin{align}
\mathcal{M}_{x} & =M_{0}\cos \left( k_{0}y\right)
\sin \left( k_{0}x\right) \sin \left( k_{0}z\right), \\
\mathcal{M}_{y} & =M_{0}\cos \left( k_{0}x\right) \sin \left( k_{0}y\right) \sin \left(
k_{0}z\right).
\end{align}
\end{subequations}
In the tight-binding approximation [see Appendix \ref{App:Tigh}], the Hamiltonian becomes
\begin{eqnarray}
H &=&-t\sum\limits_{\left\langle i,j\right\rangle }\left( \hat{c}%
_{i\uparrow }^{\dag }\hat{c}_{j\uparrow }-\hat{c}_{i\downarrow }^{\dag }\hat{%
c}_{j\downarrow }\right) +\sum\limits_{i}h_{z}\left( \hat{n}_{i\uparrow }-%
\hat{n}_{i\downarrow }\right) \notag \\
&+&\left[ \sum\limits_{j_{x}}it_{\mathrm{so}}\left( \hat{c}_{j_{x}\uparrow }^{\dag }%
\hat{c}_{j_{x}+1\downarrow }-\hat{c}_{j_{x}\uparrow }^{\dag }\hat{c}%
_{j_{x}-1\downarrow }\right) +\mathrm{H.c.}\right]  \notag \\
&+&\left[ \sum\limits_{j_{y}}t_{\mathrm{so}}\left( \hat{c}_{j_{y}\uparrow }^{\dag }%
\hat{c}_{j_{y}+1\downarrow }\emph{}-\hat{c}_{j_{y}\uparrow }^{\dag }\hat{c}%
_{j_{y}-1\downarrow }\right) +\mathrm{H.c.}\right]\!\!,
\end{eqnarray}%
where the parameters $t$ and $t_{\mathrm{so}}$ denote the spin-conserved and spin-flip hopping as shown in Figs.~1(a) and 1(b), respectively.

\begin{figure*}[tbp]
\centerline{\includegraphics[width=0.94\textwidth,clip=]{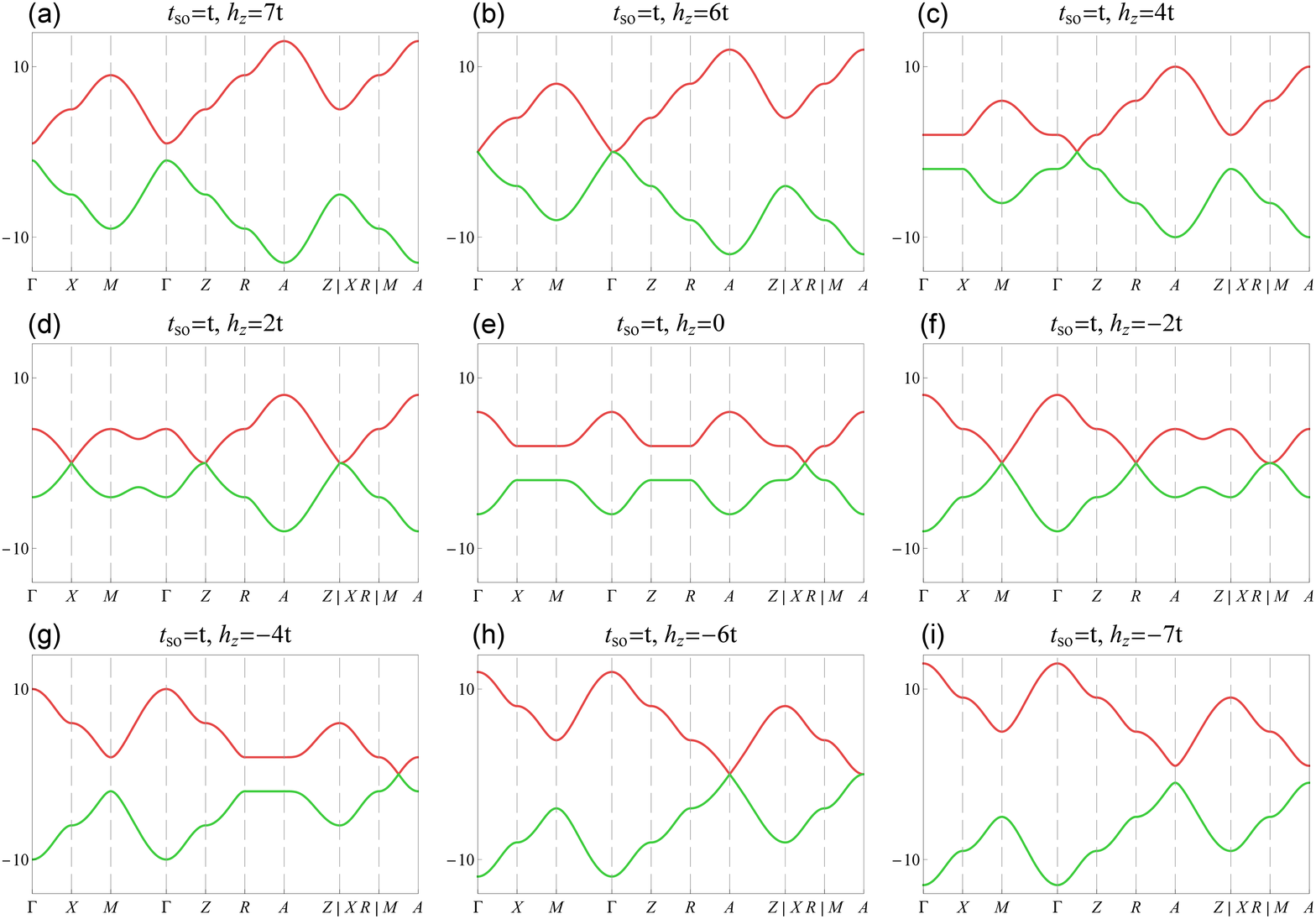}}
\caption{Band structure as a function of the Zeeman field. By adjusting the effective Zeeman field, one can observe the movement of the Weyl points in the Brillouin zones, as well as the creation and annihilation of Weyl fermions in the low-energy quasiparticle excitations. The Weyl fermions emerge at the $\Gamma-Z$, $X-R$ and $M-A$ axis as shown in Figs.~2(c), 2(e) and 2(g), but annihilate in pair at the high-symmetry points $\Gamma$, $Z$, $R$ and $A$ and become 2D massless Dirac fermions as shown in Figs.~2(b), 2(d), 2(f) and 2(h). In the case of gap opened, massive 3D Dirac fermions emerge around the high-symmetry $\Gamma$ or $A$ point with the minimum gap as shown in Figs.~2(a) and 2(i).}\label{fig2}
\end{figure*}

While the hopping amplitude is $t$ for the spin-up atoms, it is $-t$ for the spin-down ones. As a result, besides the normal Rashba type spin-orbit coupling in the x-y plane, the 3D periodic Raman potentials $\mathcal{M}_{x}$ and $\mathcal{M}_{y}$ also induce an spin-dependent hopping along all the three dimensions. This is equivalent to 3D spin-orbit coupling plus a spin-dependent hopping along the $x$ and $y$ directions. Next, we demonstrate that this Hamiltonian allows the existence of Weyl points in the single-particle energy band.

\section{Energy band structure}
Using the Fourier transformation
\begin{eqnarray}
\hat{c}_{j\sigma }=\frac{1}{\sqrt{N}}\sum\limits_{\mathbf{k}}\hat{c}_{%
\mathbf{k}\sigma }e^{i\mathbf{k\cdot r}_{j}},
\end{eqnarray}
one can obtain the Bloch Hamiltonian in the momentum space
\begin{eqnarray}
H(\mathbf{k})=d_{x}\left( \mathbf{k}\right) \sigma _{x}+d_{y}\left( \mathbf{k}\right)
\sigma _{y}+d_{z}\left( \mathbf{k}\right) \sigma _{z},
\end{eqnarray}
where the Bloch vector $\mathbf{d}(\mathbf{k})=(d_x,d_y,d_z)$ is given by
\begin{subequations}
\label{Blochvector}
\begin{align}
d_{x} & =-2t_{\mathrm{so}}\sin \left( k_{x}\right), \\
d_{y} & =-2t_{\mathrm{so}}\sin \left( k_{y}\right), \\
d_{z} & =h_{z}-2t\sum\limits_{\eta} \cos \left( k_{\eta}\right).
\end{align}
\end{subequations}
By diagonalizing the Bloch Hamiltonian, one obtains the single-particle energy bands
\begin{eqnarray}
E_{\pm}(\mathbf{k})=\pm\sqrt{d_{x}^2+d_{y}^2+d_{z}^2}.\label{band}
\end{eqnarray}
\begin{figure*}[tbp]
\centerline{\includegraphics[width=0.7\textwidth,clip=]{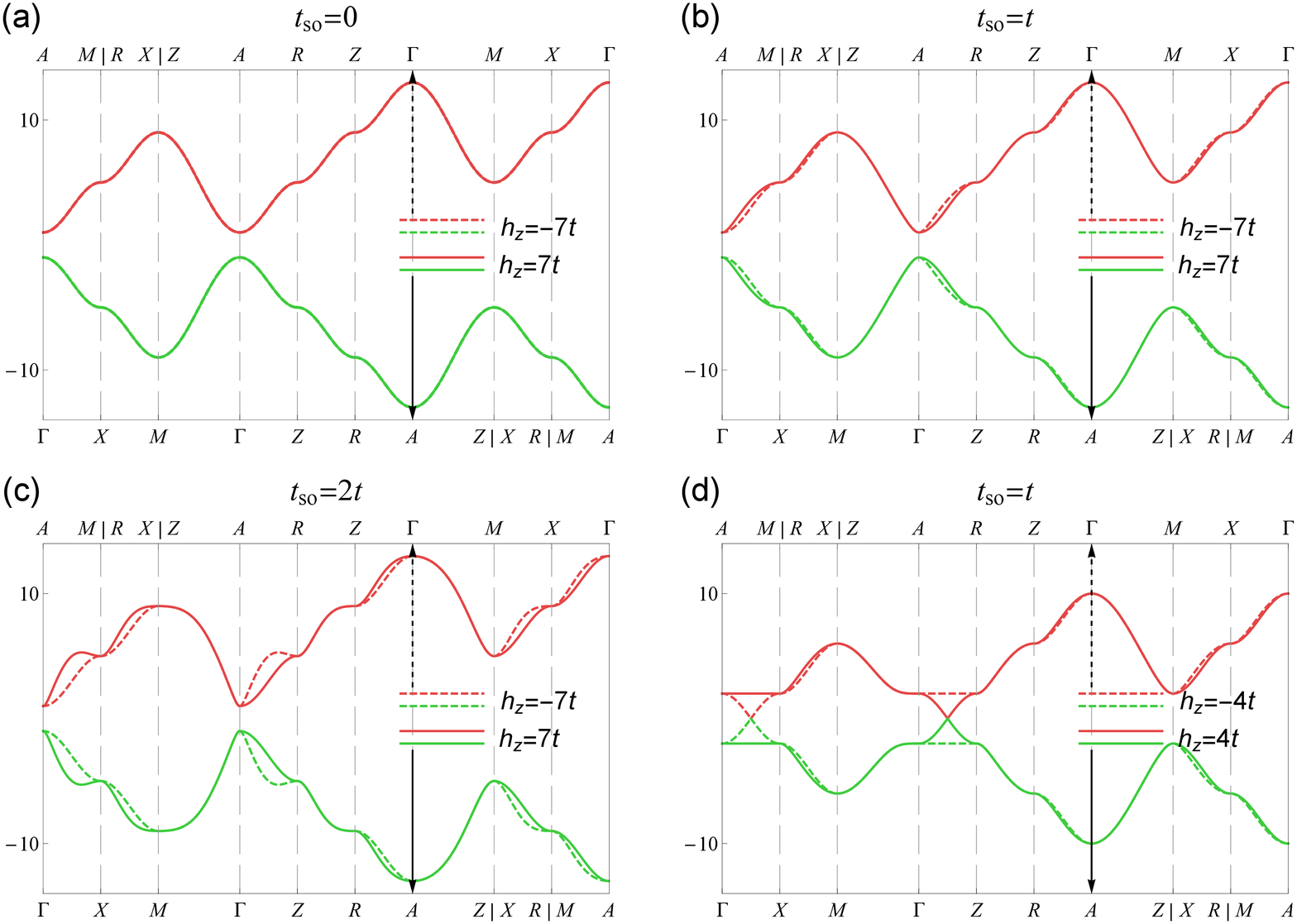}}
\caption{Explicit symmetry breaking of the band induced by spin-orbit coupling. While the solid lines represent the band on the path $\Gamma$--$X$--$M$--$\Gamma$--$Z$--$R$--$A$--$Z|X$--$R|M$--$A$, the dashed lines represent that on $A$--$M|R$--$X|Z$--$A$--$R$--$Z$--$\Gamma$--$M$--$X$--$\Gamma$ with an equal but opposite Zeeman field. The solid and dashed arrows point to the paths for solid and dashed lines of the bands respectively.}\label{fig3}
\end{figure*}

According to the high-symmetry points in the Brillouin zone as shown in Fig.~1(c), we plot the energy band structure on the path $\Gamma$--$X$--$M$--$\Gamma$--$Z$--$R$--$A$--$Z|X$--$R|M$--$A$. The band as a function of the Zeeman field is shown in Fig.~2. One can observe the opening and closing of the gap with the change of the strength of Zeeman field. When $-6t<h_z<6t$, there exists energy crossings in the momentum space. These energy crossings are actually relativistic Weyl or Dirac points, which will be demonstrated in following sections. By adjusting the effective Zeeman field, one can observe the movement of the band-crossing points in the Brillouin zones, as well as the creation and annihilation of relativistic particles in the low-energy quasiparticle excitations.

It is found that the symmetry of the energy band is explicitly broken by the spin-orbit coupling. In the absence of spin-orbit coupling, the band structures of $h_{z}$ and $-h_{z}$ on the chosen path are the same but reversed from left to right, as shown in Fig.~3(a). While in the presence of spin-orbit coupling, although the band structures of $h_{z}$ and $-h_{z}$ with reversed paths look similar, they are actually inconsistent with each other, and their difference increases with the strength of spin-orbit coupling, as shown in Figs.~3(b) and 3(c).

This can be understood from the effective expression of the bands on the chosen paths. According to Eq.~\ref{band}, the energy band on the path $\Gamma$--$X$--$M$--$\Gamma$--$Z$--$R$--$A$--$Z|X$--$R|M$--$A$ can be written as the piecewise function
\begin{subequations}
\label{band1}
\begin{align}
E_{\Gamma -X} & =\pm\sqrt{4t_{\text{so}}^{2}\sin ^{2}q+(h_{z}-4t-2t\cos q)^{2}}, \\
E_{X-M} & =\pm\sqrt{4t_{\text{so}}^{2}\sin ^{2}q+(h_{z}-2t\cos q)^{2}}, \\
E_{M-\Gamma } & =\pm\sqrt{8t_{\text{so}}^{2}\sin ^{2}\frac{\sqrt{2}q}{2}+\Big(h_{z}-2t+4t\cos \frac{\sqrt{2}q}{2}\Big)^{2}}, \\
E_{\Gamma -Z} & =\pm\sqrt{(h_{z}-4t-2t\cos q)^{2}}, \\
E_{Z-R} & =\pm\sqrt{4t_{\text{so}}^{2}\sin ^{2}q+(h_{z}-2t\cos q)^{2}}, \\
E_{R-A} & =\pm\sqrt{4t_{\text{so}}^{2}\sin ^{2}q+(h_{z}+4t-2t\cos q)^{2}}, \\
E_{A-Z} & =\pm\sqrt{8t_{\text{so}}^{2}\sin ^{2}\frac{\sqrt{2}q}{2}+\Big(h_{z}+2t+4t\cos \frac{\sqrt{2}q}{2}\Big)^{2}}, \\
E_{X-R} & =\pm\sqrt{(h_{z}-2t\cos q)^{2}}, \\
E_{M-A} & =\pm\sqrt{(h_{z}+4t-2t\cos q)^{2}}.
\end{align}
\end{subequations}
In comparison, the band on the path $A$--$M|R$--$X|Z$--$A$--$R$--$Z$--$\Gamma$--$M$--$X$--$\Gamma$ is written as
\begin{subequations}
\label{band2}
\begin{align}
E_{A-M} & =\pm\sqrt{(h_{z}+4t+2t\cos q)^{2}}, \\
E_{R-X} & =\pm\sqrt{\left( h_{z}+2t\cos q\right) ^{2}}, \\
E_{Z-A} & =\pm\sqrt{8t_{\text{so}}^{2}\sin ^{2}\frac{\sqrt{2}q}{2}+\Big(h_{z}+2t-4t\cos \frac{\sqrt{2}q}{2}\Big)^{2}}, \\
E_{A-R} & =\pm\sqrt{4t_{\text{so}}^{2}\sin ^{2}q+(h_{z}+4t+2t\cos q)^{2}}, \\
E_{R-Z} & =\pm\sqrt{4t_{\text{so}}^{2}\sin ^{2}q+(h_{z}+2t\cos q)^{2}}, \\
E_{Z-\Gamma } & =\pm\sqrt{(h_{z}-4t+2t\cos q)^{2}}, \\
E_{\Gamma -M} & =\pm\sqrt{8t_{\text{so}}^{2}\sin ^{2}\frac{\sqrt{2}q}{2}+\Big(h_{z}-2t-4t\cos \frac{\sqrt{2}q}{2}\Big)^{2}}, \\
E_{M-X} & =\pm\sqrt{4t_{\text{so}}^{2}\sin ^{2}q+(h_{z}+2t\cos q)^{2}}, \\
E_{X-\Gamma } & =\pm\sqrt{4t_{\text{so}}^{2}\sin ^{2}q+(h_{z}-4t+2t\cos q)^{2}}.
\end{align}
\end{subequations}
Here $q\in [0,\pi]$ represents the relative distance from the starting point of each piecewise function. Take $E_{\Gamma -X}$ and $E_{A-M}$ as an example. Without spin-orbit coupling, they are the same with opposite Zeeman fields. But with spin-orbit coupling, they are unequal and the difference becomes extremely obvious for a special case with $t_{\mathrm{so}}=t$ and $|h_z|=4t$, as shown in Fig.~3(d). In this case, $E_{\Gamma -X}=\pm 2|t|$ but $E_{A-M}=\pm 2|t\cos q|$.
\begin{figure}[tbp]
\centerline{\includegraphics[width=0.45\textwidth,clip=]{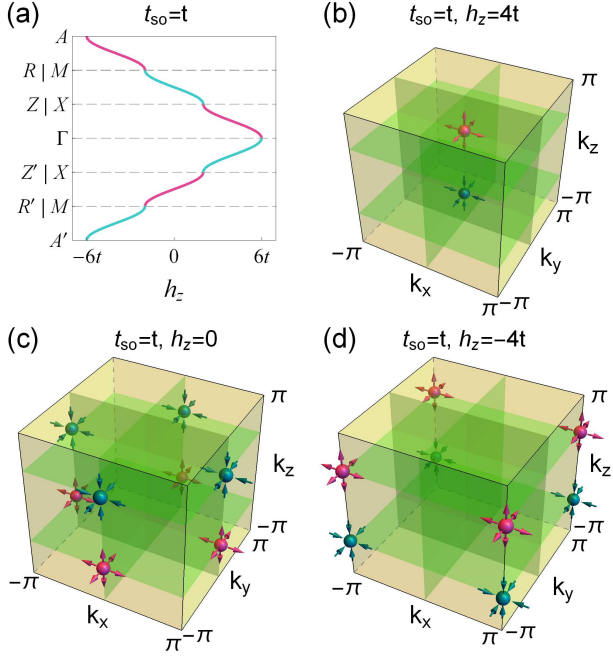}}
\caption{Position and distribution of the Weyl points. (a) The position of the Weyl points as a function of the strength of Zeeman field. The ``pink" and ``cyan" lines describe the trajectories of the
right-handed and left-handed Weyl points, respectively. (b)-(d) Three typical distributions of Weyl points in the first Brillouin zones. The Weyl points are located at (b) $(0,0,\pm \frac{\pi}{2})$ for $h_z=4$, (c) $(\pi,0,\pm \frac{\pi}{2})$ and $(0,\pi,\pm \frac{\pi}{2})$ for $h_z=0$, and (d) $(\pi,\pi,\pm \frac{\pi}{2})$ for $h_z=-4$. The Weyl points with right-handed (``pink sphere") and left-handed (``cyan sphere") chirality always arise in pairs at separated momenta, and act as the monopole and anti-monopole in the bulk Brillouin zones.}\label{fig4}
\end{figure}

\section{Simulation of Weyl fermions}
By adjusting the strength of the Zeeman fields, one can observe the movement of the Weyl points in the Brillouin zone. In Fig.~4(a), we show the position of the Weyl points in the momentum space as a function of the strength of Zeeman field. In the interval of $2t<h_z<6t$, a pair of Weyl points with opposite chirality emerge at the $\Gamma$--$Z$ axis of the Brillouin zone with $\mathbf{k}_0=(0,0,\pm \arccos{\frac{h_z-4t}{2t}})$, as shown in Fig.~4(b). In the interval of $-2t<h_z<2t$, two pairs of Weyl points emerge at the $X$--$R$ axis of the Brillouin zone with $\mathbf{k}_0=(0,\pi,\pm \arccos{\frac{h_z}{2t}})$ and $\mathbf{k}_0=(\pi,0,\pm \arccos{\frac{h_z}{2t}})$, as shown in Fig.~4(c). In the interval of $-6t<h_z<-2t$, one pair of Weyl points emerge at the $M$--$A$ axis of the Brillouin zone with $\mathbf{k}_0=(\pi,0,\pm \arccos{\frac{h_z+4t}{2t}})$, as shown in Fig.~4(d).

The trajectories of the Weyl points exactly distinguish different topological phases in this system. The corresponding phase diagram can be obtained by calculating the Chern number
\begin{eqnarray}
C=\frac{1}{4\pi}\int \hat{d}(\mathbf{k})\cdot[\partial_{k_x}\hat{d}(\mathbf{k})\times \partial_{k_y}\hat{d}(\mathbf{k})]dk_xdk_y,
\end{eqnarray}
with $\hat{d}(\mathbf{k})\equiv \mathbf{d}(\mathbf{k})/|\mathbf{d}(\mathbf{k})|$. In Fig.~5, we plot the topological phase diagram as a function of $h_z$ and $k_z$, and the trajectories of the Weyl points with right-handed and left-handed chirality are shown by ``pink" and ``cyan" lines, respectively. It is found that there exist two different topological phases characterized by $C=1$ and $C=-1$, and the upper and lower branches always carry opposite topological numbers with each other. By adjusting the effective Zeeman field $h_z$, one can observe the topological phase transition from $C=1$ to $C=-1$, then to a topological trivial phase with $C=0$.
\begin{figure}[tbp]
\centerline{\includegraphics[width=0.45\textwidth,clip=]{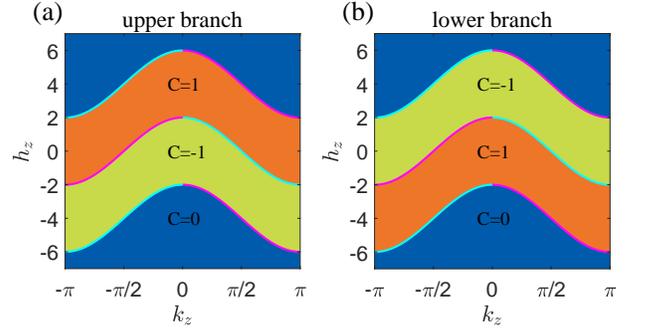}}
\caption{Topological phase diagram. The topological phase diagram as a function of the effective Zeeman strength $h_z$ and the momentum $k_z$ is plotted by calculating the Chern number $C$. There exist two different kinds of topological phases characterized by $C=1$ and $C=-1$. The trajectories of the Weyl points with right-handed and left-handed chirality are shown by ``pink" and ``cyan" lines, respectively, which exactly distinguish different topological phases in each branch.}\label{fig5}
\end{figure}

Next, we demonstrate that the dispersion relation around the band touching points is linear, and it manifests relativistic Weyl fermions in the low-energy excitations. Around the half filling of atoms in the sites of the lattice, one can expand the momentum $\mathbf{k}$ near the touching point $\mathbf{k}_0=(k_x^0,k_y^0,k_z^0)$ as $\mathbf{k}=(k_x^0+q_x,k_y^0+q_y,k_z^0+q_z)$. By approximating $H(\mathbf{k})\approx H(\mathbf{k}_0+\mathbf{q})$, one can obtain the effective Weyl Hamiltonian in the low-energy excitations
\begin{eqnarray}
H(\mathbf{q})=\upsilon_x q_x\sigma_x+\upsilon_y q_y\sigma_y+\upsilon_z q_z\sigma_z,\label{Weyl Hamiltonian}
\end{eqnarray}
where $\upsilon_x=-2t_\mathrm{so} \cos{(k_x^0)}$, $\upsilon_y=-2t_\mathrm{so} \cos{(k_y^0)}$ and $\upsilon_z=2t \sin{(k_z^0)}$. Then the energy band is given by $E_\pm(\mathbf{q})=\pm\sqrt{\upsilon_x^2q_x^2+\upsilon_y^2q_y^2+\upsilon_z^2q_z^2}$, which manifests linear dispersion for $\upsilon_x=\upsilon_y=\upsilon_z$. Without loss of generality, we consider the case with $t_\mathrm{so}=t=1$ and $h_z=4$. In this case, the Weyl points are located at $\mathbf{k}_0=(0,0,\pm\frac{\pi}{2})$. From Eq.~(\ref{Weyl Hamiltonian}), one can calculate the Berry curvature as~\cite{QNiu2010}
\begin{eqnarray}
\mathbf{\Omega}=\pm\frac{1}{2}\frac{\mathbf{q}}{|\mathbf{q}|^3}.
\end{eqnarray}
This Berry curvature manifests as a field generated by a monopole at the origin $\mathbf{q}=0$~\cite{Dirac1931,CNYang,SakuraiJ}. It should be indicated that this monopole is represented in the momentum space~\cite{ZFang}, which is essentially different from those observed with ultracold atoms in the real space~\cite{MWRay}. The topological charge of the Weyl point can be characterized by the first Chern number
\begin{eqnarray}
C=\frac{1}{2\pi}\oint_S \mathbf{\Omega}\cdot d\mathbf{S}=\pm 1
\end{eqnarray}
through any surface $S$ enclosing the points. Here the sign of the topological charge is ``$+$" for $\mathbf{k}_0=(0,0,\frac{\pi}{2})$ and ``$-$" for $\mathbf{k}_0=(0,0,-\frac{\pi}{2})$. This implies that the Weyl points always exist in pairs with opposite chirality, which is consistent with the Nielsen-Ninomiya theorem~\cite{HBNielsen}. The Weyl point with topological charge ($+1$) or ($-1$) allows the momentum and spin of the low-energy quasi-particle excitations parallel or antiparallel, and leads to right-handed or left-handed fermions. In Figs.~4(b)-4(d), we show
three representative distributions of the Weyl points in the First Brillouin zone, where the right- and left-handed Weyl points are distinguished with ``pink" and ``cyan" colors, respectively.

For some special strengths of the Zeeman field, we find that the Weyl points annihilate in pairs at the high-symmetry $\mathbf{k}$ points in the Brillouin zones. For example, when the strength of the Zeeman field is $6t$, $2t$, $-2t$ and $-6t$, the Weyl points annihilate at the high-symmetry points $\Gamma$, $Z$, $R$ and $A$, respectively [see Fig.~1(c) and Fig.~2]. Around these points, the low-energy effective Hamiltonian becomes
\begin{eqnarray}
H(\mathbf{q})=\upsilon_x q_x\sigma_x+\upsilon_y q_y\sigma_y\label{massless Dirac Hamiltonian}
\end{eqnarray}
indicating that the low-energy quasi-particle excitations are 2D massless Dirac fermions~\cite{LMDuan}. From this point, one can control the transformation of the low-energy quasi-particle excitations between 3D Weyl fermions and 2D massless Dirac fermions.

When the strength of the Zeeman field exceeds a critical value with $|h_z|>6t$, a gap with
$\Delta=|h_z|-6t$ will be opened between the two branches of the band. While the minimum gap locates at the high-symmetry $\Gamma$ point for $h_z>6t$, it locates at the high-symmetry $A$ point for $h_z<-6t$. Around the points with the minimum gap, the energy band can be approximately expressed as $E(\mathbf{q})=\pm \sqrt{\Delta^2+\upsilon_x^2 q_x^2+\upsilon_y^2 q_y^2+\upsilon_z^2 q_z^2}$ with $\upsilon_x=\upsilon_y=\sqrt{4t_s^2+2\Delta t}$ and $\upsilon_z=\sqrt{2\Delta t}$, which represents the standard energy-momentum relation for the relativistic massive 3D Dirac fermions~\cite{ABermudez2010,LMazza2012,SLZhu2010,LLepori2010}. In this case, the low-energy quasiparticle excitations around the half filling are then exactly described by the Dirac Hamiltonian
\begin{eqnarray}
H(\mathbf{q})&=&\left(
\begin{array}{cc}
0 & \sigma _{x} \\
\sigma _{x} & 0%
\end{array}%
\right) \upsilon _{x}q_{x}+\left(
\begin{array}{cc}
0 & \sigma _{y} \\
\sigma _{y} & 0%
\end{array}%
\right) \upsilon _{y}q_{y}\notag \\
&& +\left(
\begin{array}{cc}
0 & \sigma _{z} \\
\sigma _{z} & 0%
\end{array}%
\right) \upsilon _{z}q_{z}+\Delta \left(
\begin{array}{cc}
I & 0 \\
0 & -I%
\end{array}%
\right),\label{massive Dirac Hamiltonian}
\end{eqnarray}
This implies that one can observe the transformation of the low-energy quasi-particle excitations between massless Weyl fermions and massive Dirac fermions by adjusting the strength of the effective Zeeman field.

\section{Experimental detection}
In the above, we have demonstrated that cold atoms in a cubic optical lattice with Raman-assisted spin-orbit coupling allow the realization of relativistic low-energy quasiparticle excitations involving Weyl fermions, 3D massive Dirac fermions and 2D massless Dirac fermions. A more important thing is how to experimentally observe these relativistic quasiparticles and the associated quantum
phase transition. In a previous work~\cite{LMDuan}, it has been suggested that both the massive and massless 2D Dirac fermions can be experimentally detected with the density profile measurement~\cite{JRAnglin,MWZwierlein,YShin}. Here we develop the method of density profile measurement to observe the Weyl fermions as well as the transformation between different relativistic quasiparticles.

\begin{figure*}[tbp]
\centerline{\includegraphics[width=0.9\textwidth,clip=]{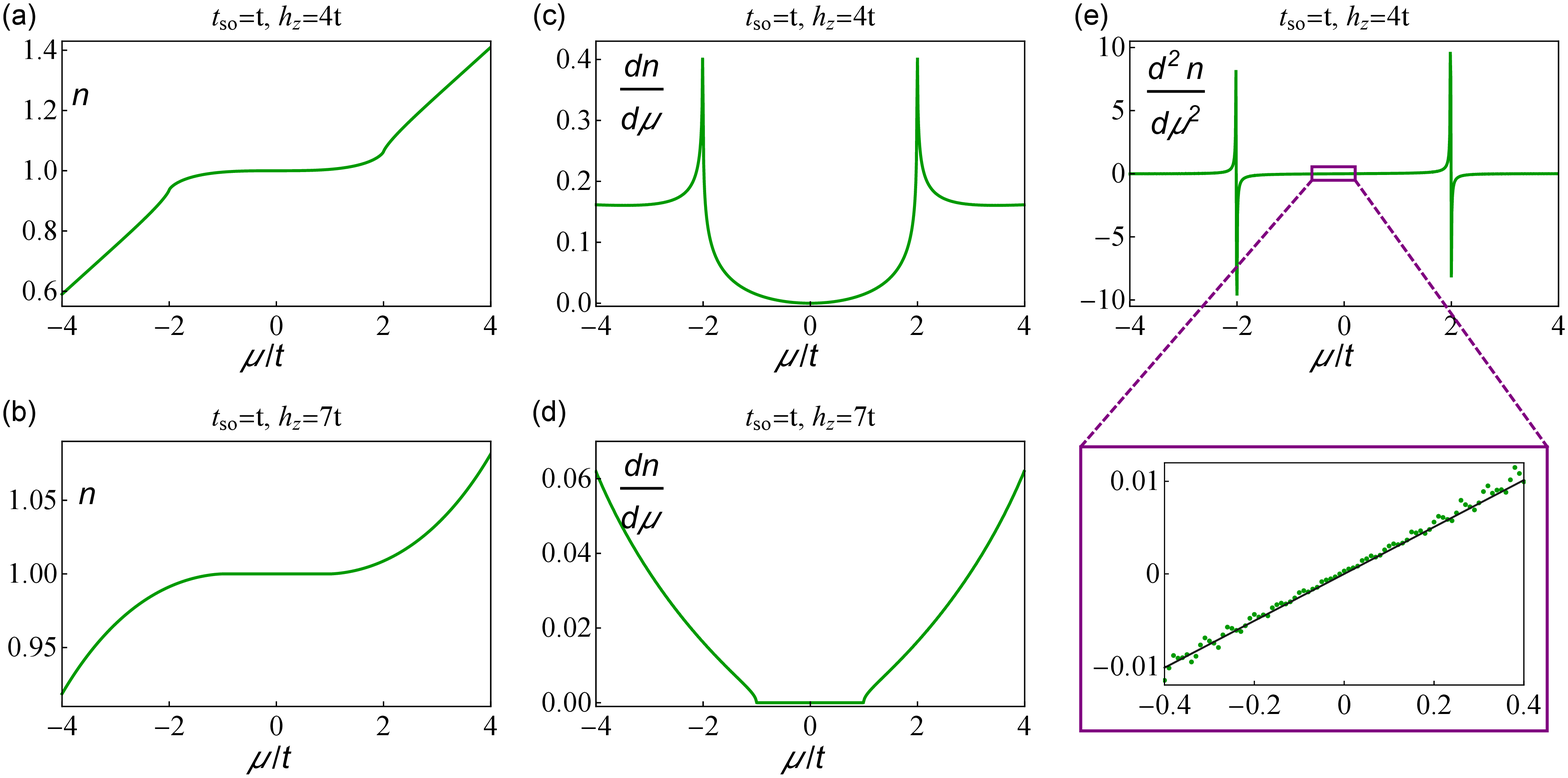}}
\caption{The density profile and its derivatives for the detection of Weyl fermions. (a)-(b) The number density of atoms $n$ per unit cell as a function of the chemical potential $\mu$ (corresponding to a rescaled atomic density profile in a trap) for (a) $h_z=4$ and (b) $h_z=7$. A plateau appears at the atom density $n(\mu)=1$ (corresponding to half filling of the band) for $h_z=7$. (c)-(d) The first derivative $\frac{dn}{d\mu}$ as a function of the chemical potential $\mu$ for (c) $h_z=4$ and (d) $h_z=7$. (e) The second derivative $\frac{d^2n}{d\mu^2}$ as a function of the chemical potential $\mu$ for $h_z=4$. The insert of Fig.~6(e) shows an enlarged part of the panel (e) at the vicinity of $\mu=0$, and the linearity of the curve shows the linear dispersion relation for the Weyl fermions in the low-energy quasiparticle excitations.}\label{fig6}
\end{figure*}

For trapped fermions, the local chemical potential can be written as $\mu=\mu_0-V(\mathbf{r})$ under local density approximation (LDA), where $\mu_0$ is the chemical
potential at the trap center and $V(\mathbf{r})=m\omega^2\mathbf{r}^2/2$ is the harmonic trapping potential~\cite{WYi2006}. This indicates that the density profile $n(\mathbf{r})$ is uniquely determined by the equation of the state $n(\mu)$. For fermions in an optical lattice, the number of atoms per unit cell can be given by
\begin{eqnarray}
n(\mu)=\frac{1}{V_0}{\int \left[\frac{1}{e^{\frac{E_+(\mathbf{k})-\mu}{T}}+1}+\!\frac{1}{e^{\frac{E_-(\mathbf{k})-\mu}{T}}+1}\right] \mathbf{d}^3\mathbf{k}},\label{state equation}
\end{eqnarray}
where $V_0=\frac{8\pi^3}{a^3}$ is the volume of the first Brillouin zone of the cubic optical lattice. In the low temperature limit with $T\sim 0$, the atomic density can be approximatively written as
\begin{eqnarray}
n(\mu)&=&\frac{1}{V_0}{\int_{E_+(\mathbf{k})\leq \mu} \mathbf{d}^3\mathbf{k}}+\frac{1}{V_0}{\int_{E_-(\mathbf{k})\leq \mu} \mathbf{d}^3\mathbf{k}}.
\end{eqnarray}
Then we have
\begin{eqnarray}
n(\mu)&=&\left\{
\begin{array}{c}
1+\frac{V}{V_{0}}\ (\mu >0), \\[3mm]
1-\frac{V}{V_{0}}\ (\mu <0),%
\end{array}%
\right.\label{state equation3}
\end{eqnarray}
where $V$ is the volume enclosed by the energy surface $E_\pm ^2(\mathbf{k})=\mu^2$.

In the case of gap opened, from Eq.~(\ref{state equation3}) one can obviously obtain $n(\mu)=1$ for $|\mu|\leq \Delta$. This implies that for the gapped phase with massive 3D Dirac fermions, there is a plateau at the atom density $n(\mu)=1$ in the density profile, as shown in Figs.~6(b) and 6(d). In contrast, for the gapless phase with Weyl fermions there is no such a plateau in the density profile, as shown in Figs.~6(a) and 6(c). This provides a crucial criterion in distinguishing the Weyl fermions and the massive 3D Dirac fermions of the low-energy quasiparticle excitations. Comparing the parameters in Fig.~6 with those of the topological phase diagram in Fig.~5, it is noted that the gapped phase with a plateau in the density profile is always topological trivial with the Chern number $C=0$. In contrast, the topological property for the gapless phase without such a plateau depends on the position in the momentum space. As the trajectories of the Weyl points exactly distinguish different topological phases as shown in Fig.~5, for the case of Figs.~6(a) and 6(c) with $h_z=4t$, when going across the Weyl point in the momentum space, one can observe the phase transition between topologically trivial and nontrivial phases.

In order to find out the evidence for the linear dispersion relation of the Weyl fermions, around the
half filling with $n=1$ (which corresponds to the touching point of the bands), we take the approximate expression of the energy band $E_\pm(\mathbf{q})=\pm\sqrt{\upsilon_x^2q_x^2+\upsilon_y^2q_y^2+\upsilon_z^2q_z^2}$. Substitute it into Eq.~(\ref{state equation3}), we have
\begin{eqnarray}
n(\mu)&=&1+\frac{4\pi\mu^3}{3V_{0}|\upsilon_x\upsilon_y\upsilon_z|}.\label{state equation4a}
\end{eqnarray}
This indicates that around the half filling, the variation $\delta n(\mu)$ of the atomic density is proportional to $(\delta\mu)^3$. As a result, the first derivative $\frac{d n}{d\mu}$ of the atomic density with respect to the chemical potential $\mu$ is a parabolic function, and the second derivative $\frac{d^2 n(\mu)}{d\mu^2}$ is linearly proportional to $\mu$ with the explicit asymptotic expression
\begin{eqnarray}
\frac{d^2 n(\mu)}{d\mu^2}=\frac{\mu}{\pi^2|\upsilon_x\upsilon_y\upsilon_z|}.\label{state equation4}
\end{eqnarray}
The linear form of the second derivative of the atomic density with respect to the chemical potential signals the linear dispersion relation around the Weyl points, which can be used to confirm the existence of massless Weyl fermions. In Figs.~6(a), 6(c) and 6(e), we show the number density of atoms $n(\mu)$ and its first and second derivatives in the phase of Weyl fermions. One can find that around the Weyl point, the results are exactly consistent with the above analysis.

For the gapless phase with massless 2D Dirac fermions at the high-symmetry $\mathbf{k}$ points, the energy band around the the touching points is independent of $k_z$, and is expressed as $E_\pm(\mathbf{q})=\pm\sqrt{\upsilon_x^2k_x^2+\upsilon_y^2k_y^2}$. When the chemical potential $\mu$ is very small, we can approximately express the number of atoms per unit cell as
\begin{eqnarray}
n(\mu)=\frac{1}{S_0}{\int \left[\frac{1}{e^{\frac{E_+(\mathbf{q})-\mu}{T}}+1}+\frac{1}{e^{\frac{E_-(\mathbf{q})-\mu}{T}}+1}\right] \mathbf{d}^2\mathbf{q}},\label{state equation666}
\end{eqnarray}
where $S_0=\frac{4\pi^2}{a^2}$ is the area of a section perpendicular to the $k_z$ axis in the first Brillouin zone. In the low temperature limit with $T\sim 0$, the atomic density can be approximatively written as
\begin{eqnarray}
n(\mu)&=&\frac{1}{S_0}{\int_{E_+(\mathbf{q})\leq \mu} \mathbf{d}^2\mathbf{q}}+\frac{1}{S_0}{\int_{E_-(\mathbf{q})\leq \mu} \mathbf{d}^2\mathbf{q}}.
\end{eqnarray}
Thus we have
\begin{eqnarray}
n(\mu)&=&\left\{
\begin{array}{c}
1+\frac{\mu^2}{4\pi|\upsilon_x\upsilon_y|}\ (\mu >0), \\[3mm]
1-\frac{\mu^2}{4\pi|\upsilon_x\upsilon_y|}\ (\mu <0).%
\end{array}%
\right.\label{state equation5}
\end{eqnarray}
Compared with Eq.~(\ref{state equation4a}), one can see that the atomic density profiles for the Weyl phase and the 2D massless Dirac phase have distinct difference. While the second derivative is linear around the Weyl points, the first derivative of the atomic density around the gapless Dirac points is linearly proportional to the chemical potential with the explicit asymptotic expression
\begin{eqnarray}
\frac{d n(\mu)}{d\mu}=\frac{|\mu|}{2\pi|\upsilon_x\upsilon_y|}.\label{state equation4}
\end{eqnarray}
This is consistent with the previous prediction on the observation of 2D massless Dirac fermions in a hexagonal optical lattice~\cite{LMDuan}.

\section{Conclusion}
In summary, we have predicted possible simulation and detection of Weyl fermions in ultracold Fermi gases with Raman-assisted spin-orbit coupling, which contributes to expanding the previous experimental observation of Weyl semimetal band in Bose gases. By changing the strength of the effective Zeeman field, we discussed the movement of the Weyl points in the Brillouin zone as well as the creation and annihilation of Weyl fermions in the low-energy quasiparticle excitations. Through explicit calculations, we showed that the atomic density profile in a trap can be used to experimentally verify the existence of Weyl fermions and the associated quantum phase transition. This provides an experimentally feasible scheme for the detection of relativistic quasiparticles in cold atom systems, which is very different from the widely used technique based on the transport measurements for condensed-matter materials. The feasibility of our proposal will motivate future experimental studies on the 3D topological orders and the observation of relativistic effects with realistic cold atom platforms.

\section*{ACKNOWLEDGMENTS}
We would like to thank Hong-Gang Luo for helpful discussions. This work was supported by the Doctoral Fund Programs of Jinzhong University under Grant No.~jzxybsjjxm2019016, the Scientific and Technological Innovation Programs of Higher Education Institutions in Shanxi under Grant No.~2020L0605, the Teaching Reform Innovation Programs of Higher Education Institutions in Shanxi under Grant No.~J2021622, and the Natural Science Basic Research Plan in Shaanxi Province of China under Grant No.~2019JQ-058.
\newline

\appendix
\renewcommand \appendixname{APPENDIX}

\section{Tight binding approximation}\label{App:Tigh}
We derive the tight-binding model from the effective Hamiltonian~\cite{XJLiu2016}
\begin{eqnarray}
H&=&\left[  \frac{\hbar^{2}\mathbf{k}^{2}}{2m}+V_{\mathrm{latt}}\left(
x,y,z\right)  \right]  \otimes\mathbf{1+}\mathcal{M}_{x}\left(  x,y,z\right)  \sigma
_{x}\notag \\
&& +\mathcal{M}_{y}\left(  x,y,z\right)  \sigma_{y}+h_{z}\sigma_{z}.\label{TBM}
\end{eqnarray}
Suppose that fermions occupy the lowest s-orbitals
$\phi_{s\sigma}\left(  \sigma=\uparrow,\downarrow\right)$, and there exists only the
nearest-neighbor hopping. The tight-binding Hamiltonian then is
given by
\begin{eqnarray}
H_{\mathrm{TI}}&=&-\sum\limits_{<i,j>,\sigma}t\hat{c}_{i\sigma}^{\dagger}\hat{c}_{j\sigma}+\sum\limits_{<i,j>}\left(  t_{\mathrm{so}}^{ij}\hat{c}_{i\uparrow}^{\dagger}\hat{c}_{j\downarrow
}+\mathrm{H.c.}\right)\notag \\
&& +\sum\limits_{i}h_{z}\left(  \hat{n}_{i\uparrow}-\hat{n}_{i\downarrow
}\right),\label{TBH}
\end{eqnarray}
where $i=(i_{x},i_{y},i_{z})$ is the 3D lattice-site index, the particle
operators $\hat{n}_{i\sigma}=\hat{c}_{i\sigma}^{\dagger}\hat{c}_{i\sigma}.$The
spin-conserved hopping coupling is induced by the lattice potential with
\begin{eqnarray}
t^{ij}\!=\!\!\int\!\! d^{3}\mathbf{r}\phi_{s\sigma}^{\left(  i\right)  }\!\left(\mathbf{r}\right)\!\left[  \frac{p_{x}^{2}+p_{y}^{2}+p_{y}^{2}}{2m}%
+V_{\mathrm{latt}}\!\left(\mathbf{r}\right)\right]\!\!\phi_{s\sigma}^{\left(
j\right)  }\!\left(\mathbf{r}\right),
\end{eqnarray}
where $V_{\mathrm{latt}}\left(  \mathbf{r}\right)=\sum\limits_{\eta}V_{\eta}\cos^{2}\left(  k_{0}\eta\right)$ and we assume $t=t^{ij}$. The spin-flip hopping coupling $t_{\mathrm{so}}^{ij}$ are driven by the Raman potentials $\mathcal{M}_{x}(x,y,z)$ and $\mathcal{M}_{y}(x,y,z)$ with
\begin{eqnarray}
t_{\mathrm{so}}^{ij}=\int d^{3}\mathbf{r}\phi_{s\uparrow}^{\left(  i\right)}\left(
\mathbf{r}\right)\left[  \mathcal{M}_{x}\sigma_{x}+\mathcal{M}_{y}\sigma
_{y}\right]\phi_{s\downarrow}^{\left(  j\right)  }\left(  \mathbf{r}\right).
\end{eqnarray}
It can be directly verified that the spin-flip hopping terms satisfy
\begin{eqnarray}
t_{\mathrm{so}}^{j_{x},j_{x}\pm1} &=&\pm i(-1)^{j_{x}+j_{y}+j_{z}}t_{\mathrm{so}},\label{SF1} \\
t_{\mathrm{so}}^{j_{y},j_{y}\pm1} &=&\pm(-1)^{j_{x}+j_{y}+j_{z}}t_{\mathrm{so}}.\label{SF2}
\end{eqnarray}

By redefining the operator $\hat{c}_{j\downarrow}\longrightarrow e^{i\pi\left(j_x+j_y+j_z\right)  }\hat{c}_{j\downarrow}$, we finally derive the Hamiltonian as
\begin{eqnarray}
H &=&\!\!-t{\displaystyle\sum\limits_{\left\langle i,j\right\rangle }}\left(\hat{c}_{i\uparrow}^{\dag}\hat{c}_{j\uparrow}-\hat{c}_{i\downarrow}^{\dag}\hat{c}_{j\downarrow}\right)  +{\displaystyle\sum\limits_{i}}h_{z}\left(  \hat{n}_{i\uparrow}-\hat{n}_{i\downarrow}\right) \notag \\
&+&\!\!\!\left[\sum\limits_{j_{x}}it_{\mathrm{so}}\!\left(  \hat{c}_{j_{x}\uparrow}^{\dag}\hat{c}_{j_{x}+1\downarrow}-\hat{c}_{j_{x}\uparrow}^{\dag}\hat{c}_{j_{x}-1\downarrow}\right)  \!+\mathrm{H.c.}\right]  \notag \\
&+&\!\!\!\left[\sum\limits_{j_{y}}t_{\mathrm{so}}\!\left(\hat{c}_{j_{y}\uparrow}^{\dag}\hat{c}_{j_{y}+1\downarrow}-\hat{c}_{j_{y}\uparrow}^{\dag}\hat{c}_{j_{y}-1\downarrow}\right)
\!+\mathrm{H.c.}\right].
\end{eqnarray}%
Transforming $H$ into momentum space yields the Bloch Hamiltonian
\begin{eqnarray}
H\left(  \mathbf{k}\right)&=&-2t_{so}%
\sin k_{x}\sigma_{x}-2t_{so}\sin k_{y}
\sigma_{y}\notag \\
&& +\left(  h_{z}-2t\sum_{\eta}\cos k_{\eta}\right)\sigma_{z}.
\end{eqnarray}


\begin{thebibliography}{99}
\bibitem{HWeyl} H. Weyl, Gravitation and the electron, Proc. Natl. Acad. Sci. USA \textbf{15}, 323 (1929).
\bibitem{XWan} X. Wan, A. M. Turner, A. Vishwanath, and S. Y. Savrasov, Topological semimetal and Fermi-arc surface states in the electronic structure of pyrochlore iridates, Phys. Rev. B \textbf{83}, 205101 (2011).
\bibitem{HWeng} H. Weng, C. Fang, Z. Fang, B. A. Bernevig, and X. Dai, Weyl Semimetal Phase in Noncentrosymmetric Transition-Metal Monophosphides, Phys. Rev. X \textbf{5}, 011029 (2015).
\bibitem{MZHasan} S.-M. Huang, S.-Y. Xu, I. Belopolski, C.-C. Lee, G. Chang, B.-K. Wang, N. Alidoust, G. Bian, M. Neupane, C. Zhang, S. Jia, A. Bansil, H. Lin, and M. Z. Hasan, A Weyl Fermion semimetal with surface Fermi arcs in the transition metal monopnictide TaAs class, Nat. Commun. \textbf{6}, 7373 (2015).
\bibitem{SYXu} S.-Y. Xu, I. Belopolski, N. Alidoust, M. Neupane, G. Bian, C. Zhang, R. Sankar, G. Chang, Z. Yuan, C.-C. Lee, S.-M. Huang, H. Zheng, J. Ma, D. S. Sanchez, B.-K. Wang, A. Bansil, F. Chou, P. P. Shibayev, H. Lin, S. Jia, and M. Z. Hasan, Discovery of a Weyl fermion semimetal and topological Fermi arcs, Science \textbf{349}, 613 (2015).
\bibitem{BQLv} B. Q. Lv, H. M. Weng, B. B. Fu, X. P. Wang, H. Miao, J. Ma, P. Richard, X. C. Huang, L. X. Zhao, G. F. Chen, Z. Fang, X. Dai, T. Qian, and H. Ding, Experimental Discovery of Weyl Semimetal TaAs, Phys. Rev. X \textbf{5}, 031013 (2015).
\bibitem{XHuang} X. Huang, L. Zhao, Y. Long, P. Wang, D. Chen, Z. Yang, H. Liang, M. Xue, H. Weng, Z. Fang, X. Dai, and G. Chen, Observation of the Chiral-Anomaly-Induced Negative Magnetoresistance in 3D Weyl Semimetal TaAs, Phys. Rev. X \textbf{5}, 031023 (2015).
\bibitem{NArmitage} N. P. Armitage, E. J. Mele, and A. Vishwanath, Weyl and Dirac semimetals in three-dimensional solids, Rev. Mod. Phys. \textbf{90}, 015001 (2018).
\bibitem{LLu} L. Lu, L. Fu, J. D. Joannopoulos, and M. Solja\v{c}i\'{c}, Weyl points and line nodes in gyroid photonic crystals, Nat. Photonics \textbf{7}, 294 (2013).
\bibitem{LLu2015} L. Lu, Z. Wang, D. Ye, L. Ran, L. Fu, J. D. Joannopoulos, and M. Solja\v{c}i\'{c}, Experimental observation of Weyl points, Science \textbf{349}, 622 (2015).
\bibitem{LLu2014} L. Lu, J. D. Joannopoulos, and M. Solja\v{c}i\'{c}, Topological photonics, Nat. Photonics \textbf{8}, 821 (2014).
\bibitem{TOzawa} T. Ozawa, H. M. Price, A. Amo, N. Goldman, M. Hafezi, L. Lu, M. C. Rechtsman, D. Schuster, J. Simon, O. Zilberberg, and I. Carusotto, Topological photonics, Rev. Mod. Phys. \textbf{91}, 015006 (2019).
\bibitem{JHJiang} J. H. Jiang, Tunable topological Weyl semimetal from simple cubic lattices with staggered fluxes, Phys. Rev. A \textbf{85}, 033640 (2012).
\bibitem{YXu} Y. Xu, R. L. Chu, and C. Zhang, Anisotropic Weyl Fermions from the Quasiparticle Excitation Spectrum of a 3D Fulde-Ferrell Superfluid, Phys. Rev. Lett. \textbf{112}, 136402 (2014); Y. Xu, F. Zhang, and C. Zhang, Structured Weyl Points in Spin-Orbit Coupled Fermionic Superfluids, Phys. Rev. Lett. \textbf{115}, 265304 (2015).
\bibitem{BLiu} B. Liu, X. Li, L. Yin, and W. V. Liu, Weyl Superfluidity in a Three-Dimensional Dipolar Fermi Gas, Phys. Rev. Lett. \textbf{114}, 045302 (2015).
\bibitem{TDubcek} T. Dub\v{c}ek, C. J. Kennedy, L. Lu, W. Ketterle, M. Solja\v{c}i\'{c}, and H. Buljan, Weyl Points in Three-Dimensional Optical Lattices: Synthetic Magnetic Monopoles in Momentum Space, Phys. Rev. Lett. \textbf{114}, 225301 (2015).
\bibitem{DWZhang} D. W. Zhang, S. L. Zhu, and Z. D. Wang, Simulating and exploring Weyl semimetal physics with cold atoms in a two-dimensional optical lattice, Phys. Rev. A \textbf{92}, 013632 (2015).
\bibitem{SGaneshan} S. Ganeshan, and S. D. Sarma, Constructing a Weyl semimetal by stacking one-dimensional topological phases, Phys. Rev. B \textbf{91}, 125438 (2015).
\bibitem{WYHe} W. Y. He, S. Zhang, and K. T. Law, Realization and detection of Weyl semimetals and the chiral anomaly in cold atomic systems, Phys. Rev. A \textbf{94}, 013606 (2016).
\bibitem{ZLi} Z. Li, H. Q. Wang, D. W. Zhang, S. L. Zhu, and D. Y. Xing, Dynamics of Weyl quasiparticles in an optical lattice, Phys. Rev. A \textbf{94} 043617 (2016).
\bibitem{YXu2} Y. Xu and L. M. Duan, Type-II Weyl points in three-dimensional cold-atom optical lattices, Phys. Rev. A \textbf{94}, 053619 (2016).
\bibitem{XKong} X. Kong, J. He, Y. Liang, and S. P. Kou, Tunable Weyl semimetal and its possible realization in optical lattices, Phys. Rev. A \textbf{95}, 033629 (2017).
\bibitem{KShastri} K. Shastri, Z. Yang, and B. Zhang, Realizing type-II Weyl points in an optical lattice, Phys. Rev. B \textbf{95}, 014306 (2017).
\bibitem{SChen2021} Z.-Y. Wang, X.-C. Cheng, B.-Z. Wang, J.-Y. Zhang, Y.-H. Lu, C.-R. Yi, S. Niu, Y. Deng, X.-J. Liu, S. Chen, and J.-W. Pan, Realization of an ideal Weyl semimetal band in a quantum gas with 3D spin-orbit coupling, Science \textbf{372}, 271 (2021).
\bibitem{XJLiu2016} Y. Q. Wang and X. J. Liu, Predicted scaling behavior of Bloch oscillation in Weyl semimetals, Phys. Rev. A \textbf{94}, 031603(R) (2016).
\bibitem{XJLiu2020} Y.-Hui Lu, B.-Z. Wang, and X.-J. Liu, Ideal Weyl semimetal with 3D spin-orbit coupled ultracold quantum gas, Sci. Bull. \textbf{65}, 2080 (2020).
\bibitem{JRAnglin} J. R. Anglin and W. Ketterle, Bose-Einstein condensation of atomic gases, Nature \textbf{416}, 211 (2002).
\bibitem{MWZwierlein} M. W. Zwierlein, A. Schirotzek, C. H. Schunck, and W. Ketterle, Fermionic superfluidity with imbalanced spin populations, Science \textbf{311}, 492 (2006); E. F. DeLong, C. M. Preston, T. Mincer, V. Rich, S. J. Hallam, N.-U. Frigaard, A. Martinez, M. B. Sullivan, R. Edwards, B. R. Brito, S. W. Chisholm, and D. M. Karl, Community genomics among stratified microbial assemblages in the ocean's interior, Science \textbf{311}, 496 (2006).
\bibitem{YShin} Y. Shin, M. W. Zwierlein, C. H. Schunck, A. Schirotzek, and W. Ketterle, Observation of Phase Separation in a Strongly Interacting Imbalanced Fermi Gas, Phys. Rev. Lett. \textbf{97}, 030401; \textbf{97}, 049901(E) (2006).
\bibitem{QNiu2010} D. Xiao, M. C. Chang, and Q. Niu, Berry phase effects on electronic properties, Rev. Mod. Phys. \textbf{82}, 1959 (2010).
\bibitem{Dirac1931} P. A. M. Dirac, Quantised singularities in the electromagnetic field. Proceedings of the Royal Society of London. Series A, \textit{Containing Papers of a Mathematical and Physical Character} \textbf{133}, 60 (1931).
\bibitem{CNYang} T. T. Wu and C. N. Yang, Concept of nonintegrable phase factors and global formulation of gauge fields, Phys. Rev. D \textbf{12}, 3845 (1975).
\bibitem{SakuraiJ} J. J. Sakurai, \textit{Modern Quantum Mechanics}, 2nd ed. (Addison Wesley, Reading, MA 1993).
\bibitem{ZFang} Z. Fang, N. Nagaosa, K. S. Takahashi, A. Asamitsu, R. Mathieu, T. Ogasawara, H. Yamada, M. Kawasaki, Y. Tokura, and K. Terakura, The anomalous Hall effect and magnetic monopoles in momentum space, Science \textbf{302}, 92 (2003).
\bibitem{MWRay} M. W. Ray, E. Ruokokoski, S. Kandel, M. M\"{o}tt\"{o}nen, and D. S. Hall, Observation of Dirac monopoles in a synthetic magnetic field, Nature \textbf{505}, 657 (2014).
\bibitem{HBNielsen} H. B. Nielsen and M. Ninomiya, Absence of neutrinos on a lattice: (I). Proof by homotopy theory, Nucl. Phys. B \textbf{185}, 20 (1981).
\bibitem{LMDuan} S. L. Zhu, B. Wang, and L. M. Duan, Simulation and Detection of Dirac Fermions with Cold Atoms in an Optical Lattice, Phys. Rev. Lett. \textbf{98}, 260402 (2007).
\bibitem{ABermudez2010} A. Bermudez, L. Mazza, M. Rizzi, N. Goldman, M. Lewenstein, and M. A. Martin-Delgado, Wilson Fermions and Axion Electrodynamics in Optical Lattices, Phys. Rev. Lett. \textbf{105}, 190404 (2010).
\bibitem{LMazza2012} L. Mazza, A. Bermudez, N. Goldman, M. Rizzi, M. A. Martin-Delgado, and M. Lewenstein, An optical-lattice-based quantum simulator for relativistic field theories and topological insulators, New J. Phys. \textbf{14}, 015007 (2012).
\bibitem{SLZhu2010} M. Yang and S. L. Zhu, Three-dimensional Dirac-like fermions in an optical lattice, Phys. Rev. A \textbf{82}, 064102 (2010).
\bibitem{LLepori2010} L. Lepori, G. Mussardo, and A. Trombettoni, (3+1) massive Dirac fermions with ultracold atoms in frustrated cubic optical lattices, Europhys. Lett. \textbf{92,} 50003 (2010).
\bibitem{WYi2006} W. Yi and L. M. Duan, Phase diagram of a polarized Fermi gas across a Feshbach resonance in a potential trap, Phys. Rev. A \textbf{74}, 013610 (2006).
\end{thebibliography}
\end{document}